# Coherent Single Photon Emission from Colloidal Lead Halide Perovskite Quantum Dots

Hendrik Utzat[1], Weiwei Sun[1], Alexander E.K. Kaplan[1], Franziska Krieg[2,3], Matthias Ginterseder[1], Boris Spokoyny[1], Nathan D. Klein[1], Katherine E. Shulenberger[1], Collin F. Perkinson[1], Maksym V. Kovalenko[2,3], and Moungi G. Bawendi[1]*

[1] Department of Chemistry, Massachusetts Institute of Technology, 77 Massachusetts Avenue Cambridge, MA 02139.

[2] Institute of Inorganic Chemistry, Department of Chemistry and Applied Bioscience, ETH Zurich, 8093 Zurich, Switzerland.

[3] Laboratory for Thin Films and Photovoltaics, Empa − Swiss Federal Laboratories for Materials Science and Technology, CH-8600 Dübendorf, Switzerland.

*Correspondence to: mgb@mit.edu

**Chemically prepared colloidal semiconductor quantum dots have long been proposed as scalable and color-tunable single emitters in quantum optics, but they have typically suffered from prohibitively incoherent emission. We now demonstrate that individual colloidal lead halide perovskite quantum dots (PQDs) display highly efficient single photon emission with optical coherence times as long as 80 ps, an appreciable fraction of their 210 ps radiative lifetimes. These measurements suggest that PQDs should be explored as building blocks in sources of indistinguishable single photons and entangled photon pairs. Our results present a starting point for the rational design of lead halide perovskite-based quantum emitters with fast emission, wide spectral-tunability, scalable production, and which benefit from the hybrid-integration with nano-photonic components that has been demonstrated for colloidal materials.**

Many proposed schemes of quantum information processing require scalable quantum emitters (QEs) capable of producing indistinguishable single photons or entangled photon pairs*(1, 2)*. To realize such QEs, the optical coherence time of the emitter ($T_2$) needs to approach twice the spontaneous emission lifetime ($2T_1$). Design of solid-state QEs with transform-limited photon coherence ($T_2 = 2T_1$) is fundamentally hampered by the exciton-bath interaction leading to optical decoherence due to phonon-scattering*(3)* and spin-noise*(4)* often resulting in $T_2 \ll 2T_1$. In addition, charge density fluctuations in the environment typically cause jumping of the spectral line of the emitter leading to further decoherence on timescales of micro- to milliseconds*(5)*. The consequence is that only a few practical QEs have been demonstrated. Atom-like defects in diamond*(6, 7)* and self-assembled III-V quantum dots (QDs) are most commonly studied, often integrated with optical micro-cavities to increase the degree of coherence in the Purcell regime*(8)*. There has also been some proof-of-concept work with single molecules*(9)*. The growth and integration of III-V QDs with high photon coherence requires the highest possible material quality and exceptional control over growth conditions. Self-assembled QDs also suffer from low scalability and typically random growth that complicates deterministic integration with micro-photonic components, shortcomings that are further aggravated by dot-to-dot inhomogeneities*(2)*. For defects in diamond, the bulky host-matrix presents a challenge for coupling to cavities and hampers efficient photon out-coupling, limiting their brightness. Conversely, chemically-synthesized colloidal semiconductor quantum dots (CQDs) exhibit unmatched ease of processability from solution, and hybrid-integration of single CQD emitters with various pre-fabricated micro-photonic components has been demonstrated. However, CQDs typically suffer from incoherent and unstable emission, which has prevented their applications in quantum optics*(1)*.

We demonstrate that a unique combination of fast radiative lifetimes and long optical coherence times gives rise to highly coherent single photon emission from individual cesium lead halide $CsPbX_3$ (X = Cl, Br, I) perovskite quantum dots (PQDs)[10, 11]. Our results suggest that – unlike any other colloidal quantum dot material – perovskite-based quantum dots can be explored as low-cost, scalable QEs with the potential for cavity-integration to generate wavelength-tunable sources of indistinguishable single photons and entangled photon pairs in the visible spectral region.

PQDs combine the advantages of chemical synthesis in large batches and precise control over the size and shape inherent to colloidal materials, with the extensive compositional tunability of lead halide perovskites. At room temperature, they display narrowband emission across the entire visible spectrum and near-unity emission quantum yields due to the remarkable defect tolerance of lead halide perovskites. The $CsPbBr_3$ PQDs used in this study were synthesized according to Ref. [11] and are stabilized with zwitterionic ligands with functional amino and sulfonate groups, which have demonstrated increased stability compared to conventional PQDs that have oleylamine and oleic acid surface ligands[10, 11]. Detailed information on the synthetic procedure can be found in the materials and methods section below. PQDs exhibit an orthorhombic (P$nma$) crystal structure consisting of corner-sharing $[PbBr_6]^-$ octahedra with $Cs^+$ ions filling the inter-octahedral voids (Fig. 1A). We confirm this structure by x-ray diffraction (Fig. S2). High resolution scanning transmission electron micrographs (Fig.1 B, C) confirm the high degree of size uniformity and a cubic quantum dot shape as reported previously[11]. The PQDs in our study have an average edge length of $13.5 \pm 2$ nm (Fig. S1). Their ensemble absorption and emission spectra are presented in Fig. 1D. The room temperature absorption edge at 2.42 eV exhibits an excitonic feature confirming quantum confinement. The emission peak centered around 2.38 eV has a full width at half maximum (FWHM) of ~ 90 meV, close to the room temperature single particle emission linewidth, confirming the high synthetic quality of our samples*(12)*.

The emission of PQDs occurs from weakly confined excitons. Due to strong spin-orbit coupling in lead halide perovskites, spin- and orbital angular momenta are strongly mixed and only the total angular moment $J$ is conserved. The exciton in PQDs is formed from a hole with s-like symmetry in the valence band and a two-fold degenerate electron in the conduction band with total angular momentum $J_{h/e} = 1/2$ *(13)*. The electron-hole exchange interaction lifts the degeneracy between the singlet ($J_{exc} = 0$) and triplet ($J_{exc} = 1$) exciton states (Fig. 1E). Due to a strong Rashba effect, the degeneracy of different angular momentum projection states $|j| = \pm 1, 0$ is additionally lifted as a result of inversion symmetry breaking in the orthorhombic crystal structure, leading to energetic splitting of the triplet excitonic fine structure with values we define as $\Omega_1$ and $\Omega_2$ *(14)*. It has recently been shown that the lowest lying triplet state in PQDs is optically bright while the singlet is dark, which is a unique feature of lead halide perovskite semiconductors*(15)*.

We spin-coated dilute solutions of PQDs on quartz to perform single-emitter characterization at low temperatures. Characterization spectra of individual PQDs with either one (A-C) or two (D) sharp emission peaks are shown in Figure 2. The insets show the degree of emission peak polarization by plotting relative transmission intensities through a linear polarizer as a function of polarizer angle. The observed linear polarization for all PQDs confirms emission from the exciton state, as the trion emission is non-polarized *(15)*. In these characterization spectra, the resolution limit of our spectrometer (~500 μeV), and potentially fast spectral diffusion, limits access to the true homogeneous linewidth and to the fine structure for PQDs with small energetic splitting between exciton sub-levels $\Omega_1(\Omega_2)$. Fig. 2E shows a typical spectral and intensity time series of a single PQD at 3.6 K under non-resonant pulsed excitation (480 nm, 50 W/cm$^2$) confirming the absence of major fluorescence intermittency or large spectral jumps, in stark contrast to established II-VI CQDs which suffer from blinking and large charge-induced spectral jumps at low temperatures*(16, 17)*. We note that some PQDs show near-Poissonian photon emission statistics over the course of minutes (Fig. S4).

A single PQD intensity correlation ($g^{(2)}(\tau)$) under pulsed excitation is shown in Fig. 2F (upper panel). Our PQDs show high biexciton emission quantum yields at low temperatures, as expected from their very high radiative rates which outcompetes any non-radiative Auger process. Their superior spectral stability compared to other CQDs allows spectral selection of the exciton ZPL and rejection of biexciton emission, analogous to single photon sources based on self-assembled QDs*(2)*. When the emission is isolated with tunable dichroic edgepass filters, PQDs exhibit strong anti-bunching ($g^{(2)}(0) < 0.04$), indicating high purity single photon emission (Fig. 2F, lower panel). All studied individual PQDs show fast photoluminescence decays that can be fit with the sum of two exponentials (Fig. 2G). The fast component (~210-270 ps) is roughly two orders of magnitude higher in intensity than the long tail, confirming their fast emission rates compared to any other single photon emitters due to the bright nature of the lowest-lying exciton ground-state and a giant oscillator strength effect*(17, 18)*. As the emission quantum yield of the PQDs used in this study is ~95% (see Fig. S8), the fast observed PL lifetime is close to the spontaneous radiative lifetime $T_1$ *(15)*. The transform-limited linewidth, calculated as $\Gamma_{rad} = \hbar/T_1$, is 3.1, 2.5, 2.4 and 2.4 µeV for PQDs 1-4, respectively.

We measure the optical coherence time and resolve the fine-structure splitting with Photon-Correlation-Fourier Spectroscopy (PCFS) *(19, 20)*, which can overcome both the temporal and spectral resolution limitations of other techniques by encoding the spectral coherence of a single emitter in intensity anti-correlations recorded at the output of a Michelson interferometer (Fig. 3A). We note that PCFS and Hong-Ou-Mandel spectroscopy, a commonly used two-photon interference method, both provide optical coherence times of single emitters on fast timescales, but PCFS allows easier extraction of spectral diffusion dynamics, ZPL fraction, and fine-structure splitting*(20)*. Unlike in conventional Fourier spectroscopy, where the interferogram is resolved by collecting a sufficient number of photons at each interferometer position (typically hundreds of milliseconds integration time), PCFS correlates photon-pairs as a function of their temporal separation $\tau$ at each interferometer position. The temporal resolution of PCFS is only determined by the photon-shot-noise at a given $\tau$, which enables measurement of the optical coherence on timescales inaccessible to most other techniques. The observable is the PCFS interferogram $G^{(2)}(\delta,\tau)$ which intuitively corresponds to the envelope of the squared interferogram compiled from photon-pairs separated by $\tau$ (Fig. 3A and notes in SI). The Fourier transform of $G^{(2)}(\delta,\tau)$ yields the spectral correlation $p(\zeta,\tau)$ defined as $p(\zeta,\tau) = \langle \int_{-\infty}^{\infty} s(\omega,t)s(\omega+\zeta,t+\tau)\,d\omega \rangle$, where $\langle \ldots \rangle$ denotes the time average and $s(\omega,t)$ represents the spectrum of the single emitter at time $t$. The spectral correlation is the sum of the auto-correlations of the spectra compiled from photon-pairs separated by $\tau$. Since the probability of spectral wandering vanishes as $\tau$ approaches zero, the spectral correlation at small $\tau$ reduces to the auto-correlation of the homogenous spectrum*(19)*.

The PCFS interferogram $G^{(2)}(\delta,\tau)$ and the corresponding spectral correlation $p(\zeta,\tau)$ (insets) are shown in Fig. 3 B-E for PQDs 1-4 at 3.6 K and for small photon lag-times of $\tau < 100$ μs where we observe the homogeneous spectral information, unaffected by spectral diffusion. For all PQDs, $G^{(2)}(\delta,\tau < 100\ \mu s)$ shows an initial fast decay due to a fast partial decoherence of the emission. The exact origin of the decay is unknown, but is likely due to a broad acoustic phonon side-band or fast population relaxation between emissive fine-structure states. The long component in $G^{(2)}(\delta,\tau)$ extending over path-length differences of $\delta \gg 1$ ps implies long optical coherence times of the ZPL emission. Importantly, the fraction of photons emitted into the coherent ZPL can be calculated as the square root of the coherent decay amplitude of $G^{(2)}(\delta)$ and ranges between ~0.5 and ~0.8, implying that the majority of photons are emitted coherently. This ZPL fraction is still smaller than for epitaxial III-V QDs, but already comparable to silicon vacancy centers in diamond (ZPL fraction 0.7), which are often studied in quantum photonics *(21)*.

With different orientations of an individual PQDs, one, two, or three emissive fine-structure states can be observed. The beatings in the interferograms arise from the energy difference between the fine-structure states modulating the envelope of the interferogram as clarified in Fig. 3A for a PQD with two observable fine-structure states. The corresponding spectral correlations $p(\zeta,\tau)$ indicate narrow lines with either three, five, or seven peaks depending on the number of observable emissive states and their splittings.

Assuming exponential dephasing (Lorentzian spectral lineshapes) for each emissive fine-structure state, we fit the data with the auto-correlation of two or three Lorentzian peaks of width $\Gamma$ defined as $\Gamma = \frac{2\hbar}{T_2}$ and energetic separations of $\Omega_i$. That is,

$$p(\zeta) = \langle \int_{-\infty}^{\infty} s(\omega)s(\omega+\zeta)\,d\omega \rangle + c$$

where $c$ is a constant accounting for comparatively broad background emission and $s(\omega)$ the spectral lineshape:

$$s(\omega) = \frac{\frac{1}{2}\Gamma}{\omega^2 - (\frac{1}{2}\Gamma)^2} + \sum_{i=1}^{n=1 \vee 2} a_i \frac{\frac{1}{2}\Gamma}{(\omega - \Omega_i)^2 - (\frac{1}{2}\Gamma)^2}$$

The decay of the optical coherence with $e^{-\frac{2}{T_2}\delta}$ and the beating patterns in the interferograms due to the fine-structure are well captured by our fit, allowing extraction of the optical coherence time $T_2$. For PQDs 1-4, we find coherence times of $T_2 \sim$ 78 ps (68, 88), 52 ps (42, 72), 50 ps (46, 54) and 66 ps (52, 92), with the confidence intervals given in parentheses. These dephasing times correspond to linewidths of $\Gamma \sim$ 16, 27, 25 and 21 µeV, respectively. Remarkably, comparison of the optical coherence times $T_2$ with the spontaneous radiative lifetimes $T_1$ of 210, 266, 272 and 269 ps shows that the PQD emission linewidth consistently approaches the transform limit ($\frac{T_2}{2T_1} \sim$ 0.19, 0.10, 0.09, 0.12 for PQDs 1-4). This value of $\frac{T_2}{2T_1}$ is two orders of magnitude higher than for the best, specially engineered II-VI hetero-structure CQDs studied to date which exhibit slow photon release from dark exciton states and small fractions of coherent photon emission*(22, 23)*. Indeed, although no synthetic optimization of the photon coherence of PQDs has been conducted, our highest value $\frac{T_2}{2T_1} \sim$ 0.2 is already comparable to $\frac{T_2}{2T_1} \sim$ 0.16 - 0.8 for typical epitaxial III-V quantum dots *(4, 8, 24, 25)*.

We further analyze the dynamics of millisecond spectral diffusion in PQDs, typically not measured even for the most established QEs. In Fig. 4A, we show the evolution of the spectral correlation $p(\zeta, \tau)$ with increasing photon lag-times $\tau$ for PQD 2. The shape progression of $p(\zeta, \tau)$ reveals that spectral diffusion in this PQD occurs via discrete spectral jumps of the homogeneous spectrum within a Gaussian envelope (Gaussian Discrete Jump Model - Fig. 4B, Fig S7), reminiscent of what has been observed for colloidal II-VI QDs*(26)*. However, the spectral jumping behavior is dramatically reduced in most PQDs as shown in Fig. 4C for PQD 3, which shows minimal broadening of the spectral correlation with increasing temporal separation of photon pairs. This minimal broadening is made apparent in Fig. 4D, where the FWHM of $p(\zeta,\tau)$ versus $\tau$ is shown. PQD 3 shows only minor broadening of the spectral correlation over the timescales investigated (30 µs to 100 ms) suggesting that QEs based on PQDs may be used to generate coherent single photons with high spectral stability. Spectral diffusion has previously been associated with fluctuating surface-charge distributions in shallow traps*(27)*. We propose that the known absence of mid-gap surface traps in lead halide perovskites*(28)* may be responsible for the resiliency of the emissive triplet from charge density fluctuation induced spectral diffusion in chemically made PQDs – even in the absence of surface passivation.

Based on these findings, we suggest that PQDs can serve as scalable building blocks in sources of quantum light with spectral-tunability over the entire visible range – a prospect hard to envision with any other quantum emitter. When integrated with optical cavities, even a very moderate Purcell enhancement of the emission rate of ~5-10 should consistently yield truly transform-limited emission and thus the emission of indistinguishable single photons. Generation of polarization entangled photon pairs via the biexciton-exciton cascade emission could potentially exhibit high efficiencies due to near unity biexciton and exciton emission quantum yields. Pursuits in this direction will benefit from the hybrid-integration of CQDs with cavities, which has been demonstrated in a multitude of pilot studies. Indeed, integration with plasmonic gap cavities*(29)*, dielectric slot waveguides*(30)*, or high-Q micro-pillar cavities*(31)* has been shown. Efficient light-cavity coupling has also been achieved in plasmonic-QD hybrid nanostructures, produced by all-chemical means, which offers a radically different approach*(32)*. In view of theses prospects, optimization of the intrinsic PQD coherence time should be explored. Tuning of the dephasing time has been demonstrated for colloidal II-VI QDs by leveraging synthetic control over the fine-structure splitting to reduce phonon-mediated dephasing*(22)*. Similar strategies may apply to PQDs, although future elucidation of the pure dephasing mechanism in single PQDs may suggest different structural handles to the dephasing time. We suspect that further passivation of the PQD surface, through suitable ligands or growth of inorganic shells, may reduce the phonon spectral density and increase the coherence time and coherent fraction of the emission. These efforts may also further reduce spectral diffusion.

Our results suggest that lead halide perovskites – with their high defect tolerance and optically bright lowest-lying exciton state – are promising semiconductors for the scalable production of quantum emitters with highly coherent emission that can be processed onto virtually any substrate and integrated with nano-photonic components. Rational optimization of these emitters will build on the tools of colloidal chemistry and the structural versatility of lead halide perovskites.

**Acknowledgments:** We gratefully acknowledge helpful discussion with Professor William Tisdale.

**Funding:** The lead authors of this study (H.U., W.S., A. K.), and N.K were funded by the United States Department of Energy, Office of Basic Energy Sciences, Division of Materials Sciences and Engineering (Award No. DE-FG02-07ER46454). K.S. was supported by the Center for Excitonics, an Energy Frontier Research Center funded by the U.S. Department of Energy, Office of Science, Basic Energy Sciences under Award No. DE-SC0001088. B.S. was funded through the Institute for Soldier Nanotechnologies. M.G. was by NSF under award EECS-1449291. CP was supported by an NSF GRFP fellowship. M.K. and F.K. acknowledge the financial support from the Swiss Federal Commission for Technology and Innovation (CTI-No. 18614.1 PFNM-NM).

**Author contributions:** H.U. conceived of the study and experiments, built the experimental setup, performed the single emitter characterization, PCFS measurements, and data modeling and interpretation. W.S and A.K. assisted with single particle spectroscopy and data analysis. B.S. and K.S. helped with developing data analysis software. F.K., M.G., C.P., and M.K. synthesized perovskite quantum dots and performed ensemble characterization (Abs, PL, QY, and TEM). N.K. developed a laser system for acquisition of fast single emitter lifetimes. H.U. and all authors interpreted the data under supervision of M.B. H.U. wrote the manuscript with input from all authors.

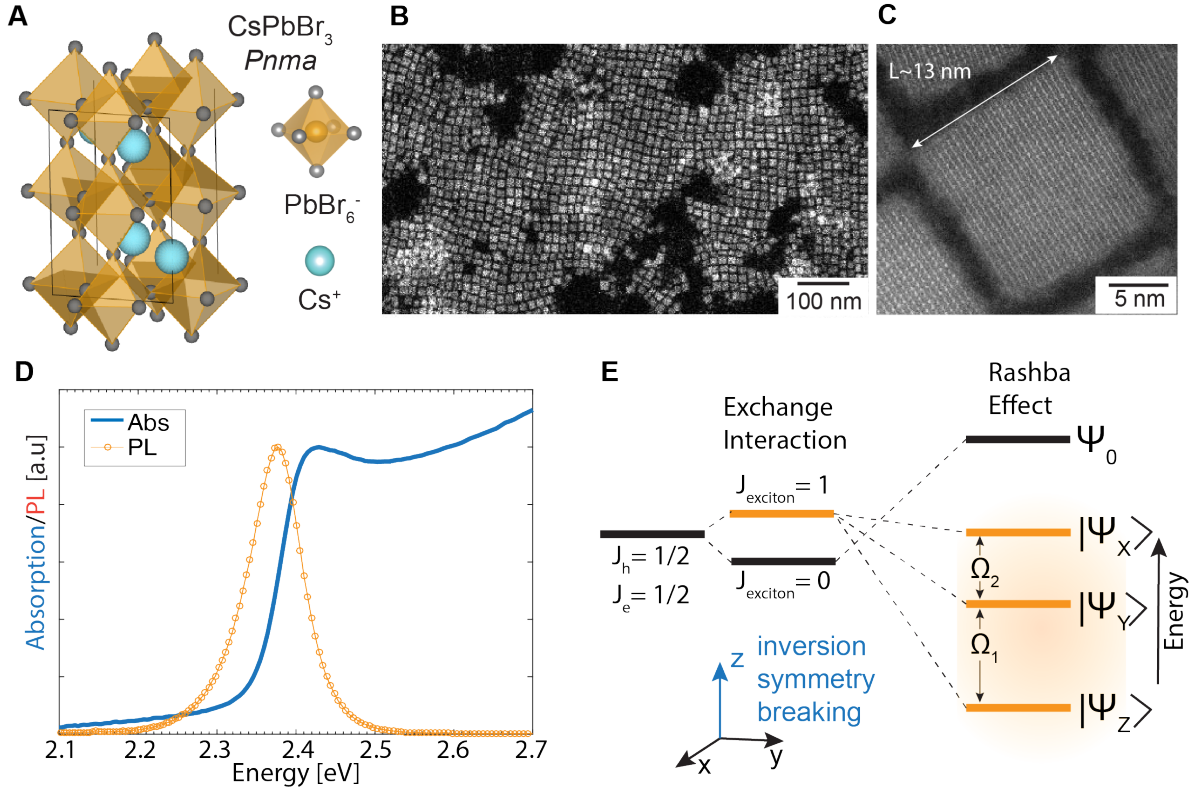

**Fig. 1.** Properties of cesium lead bromide PQDs. The Perovskite crystal structure is formed from [PbBr$_6$]$^-$ units as corner-connected octahedra, the cesium atoms occupy the voids in between (**A**). High-resolution TEM pictures confirm the cubic shape with an average edge length of ~ 13 nm and high degree of size uniformity at scale (**B, C**). Room temperature ensemble absorption and emission spectra of the PQDs studied (**D**). The emission energy was 2.38 eV with a FWHM of ~ 90 meV, indicating minimal inhomogeneous broadening. The emission of PQDs exhibits an excitonic fine-structure (**E**). Due to strong spin-orbit coupling, the total angular momentum of the electron and the hole are good quantum numbers ($J_e = J_h = 1/2$). Exchange interaction splits the exciton into a singlet ($J = 0$) and a triplet ($J = 1$). The triplet state is split further according to its angular moment projections due to the Rashba effect. For PQDs with orthorhombic Bravais lattice, the $|j| = \pm 1, 0$ degeneracy is lifted. The triplet states are emissive, while the singlet is optically dark.

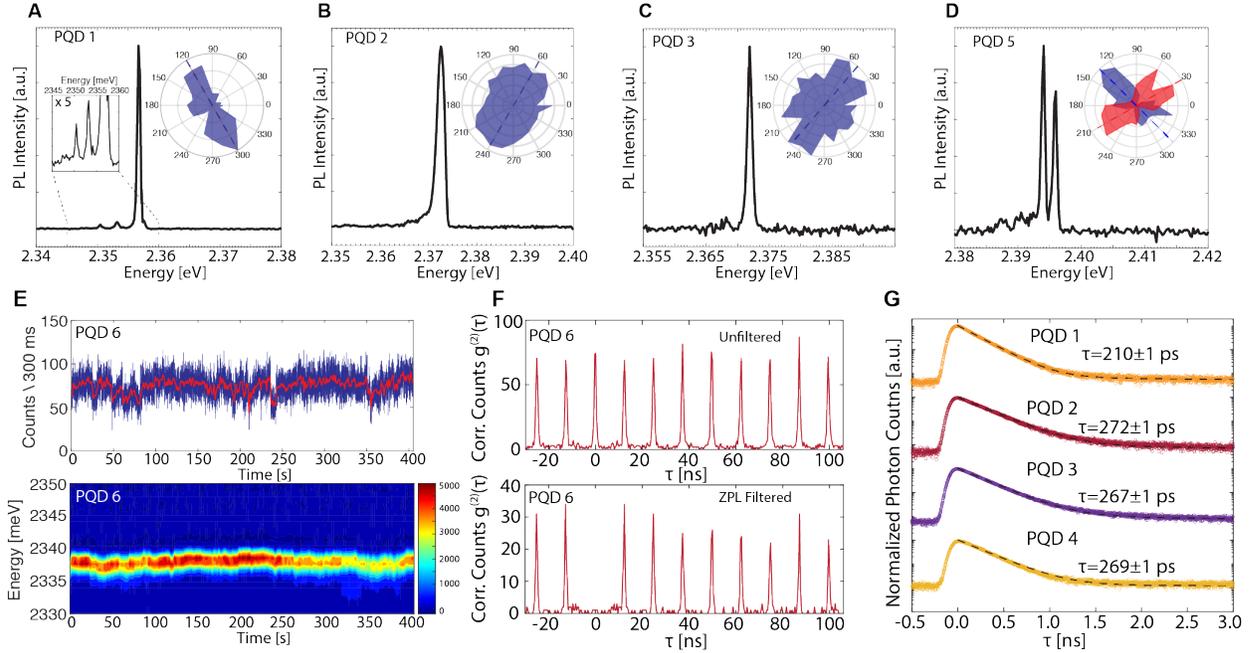

**Fig. 2.** Single PQD characterization at 3.6 K. Single PQD spectra indicate single lines with two TO phonon sidebands on the lower energy side. The resolution limit of our spectrometer does not permit resolution of the fine-structure underlying each line. The insets show the polarization dependent relative emission intensity (**A, B, C**). The emission spectrum of a typical PQD displaying two emission lines with orthogonal polarization and large enough fine-structure splitting to be resolved with the spectrometer is shown in (**D**). Single PQDs show stable emission with minimal spectral and intensity fluctuations over the course of minutes (**E**). Single PQDs exhibit high biexciton emission quantum yields as seen in the second-order intensity correlation $g^{(2)}(\tau)$ (**F, upper panel**). When the ZPL is spectrally selected sub-Poissonian emission with high single photon purity is observed (**F, lower panel**). The emission lifetime of single PQD is fast (~210 - 280 ps) and follows a mono-exponential decay over two orders of magnitude. A residual long-lived component can be observed at longer timescales, likely due to partial dark exciton emission or trapping and recapturing (**G**).

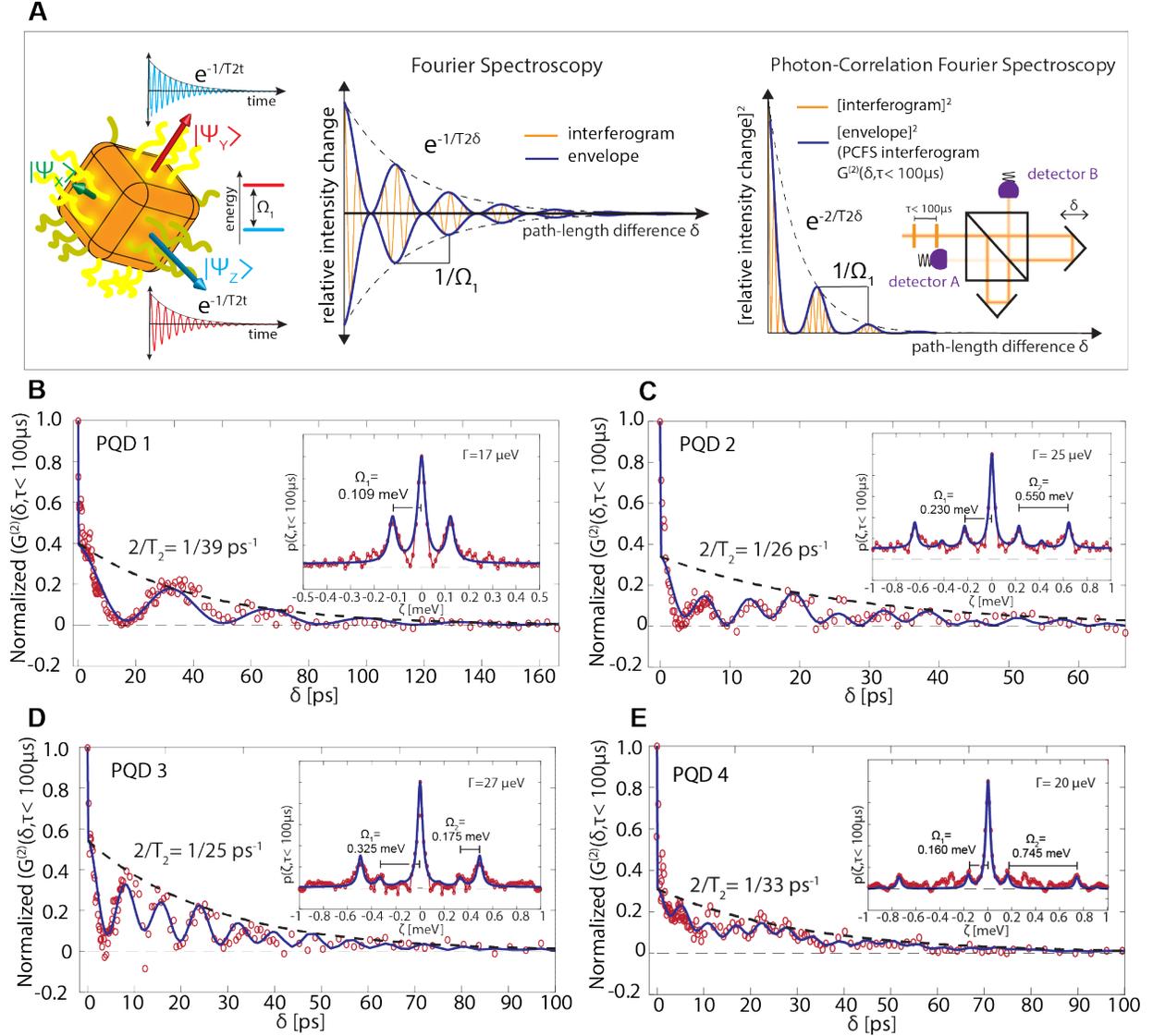

**Fig. 3.** Measurements of the optical coherence times of single PQDs. The PCFS experiment for a PQD with two observable fine-structure states of energetic splitting $\Omega_1$ is depicted in (**A**). The interferogram shows modulation of the envelope with a frequency corresponding to the energy different between the two fine-structure states $|\psi_Y\rangle$ and $|\psi_Z\rangle$ and loss of photon coherence decaying with $e^{-1/T_2 t}$. PCFS measures the envelope of the interferogram squared, compiled from photons with small temporal separation. The data for the PQDs and the corresponding spectral correlation $p(\zeta, \tau)$ (insets) for short inter-photon arrival times ($\tau < 100$ μs) where the effect of spectral diffusion is minimal are shown in (**B,C,D,E**). The blue line shows the best fit with our lineshape model, the black dashed lines indicate the exponential dephasing component of the PCFS interferogram decaying with $e^{-2/T_2 \delta}$. A fast partial loss of coherence to ~ 0.3 - 0.6

of the initial amplitude – possibly due to a broad acoustic feature or fast relaxation within the fine-structure – can be observed in the interferograms. We extract long optical coherence times of $T_2 \sim$ 50-78 ps. The widths of the underlying Lorentzian lines $\Gamma_{hom} = \frac{2\hbar}{T_2}$ and the fine-structure splittings $\Omega_1(\Omega_2)$ are indicated in the plot of the spectral correlations. For PQD 4, an additional unknown side-peak is observed that is not captured by our model, likely due to aliasing common in Fourier spectroscopy.

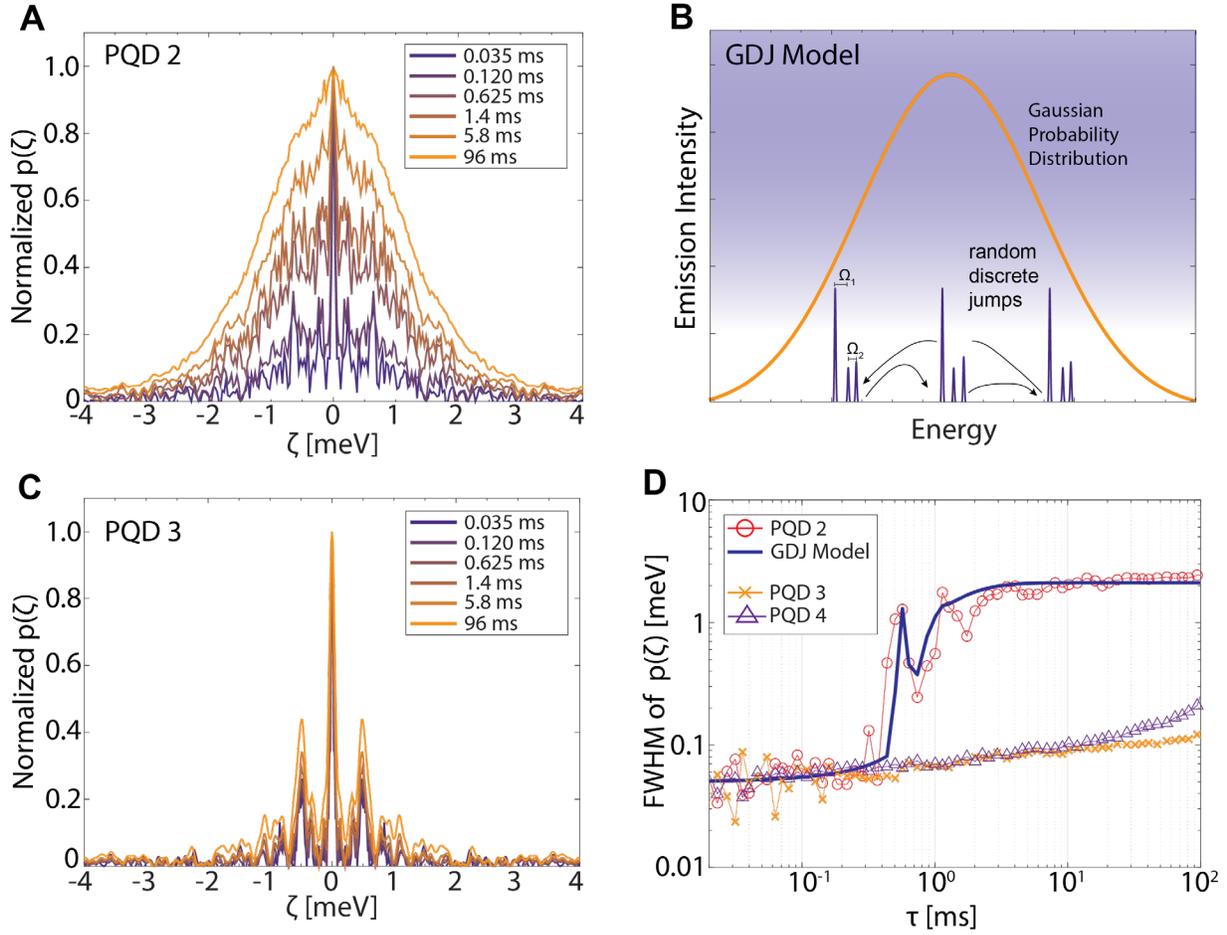

**Fig. 4.** Analysis of the spectral diffusion behavior of PQDs. The spectral correlation of PQD 2 broadens significantly with increasing photon-lag time indicating spectral diffusion (**A**). The shape evolution of $p(\zeta, \tau)$ demonstrates that spectral diffusion in PQDs is well described by a Gaussian discrete jumping model (GDJ) (**B**) in which the homogeneous spectrum undergoes rapid and discrete jumping within a Gaussian envelope. Most PQDs show drastically reduced spectral diffusion as shown in (**c**) for PQD 3. Panel (**D**) compares the FWHM of $p(\zeta, \tau)$ versus $\tau$ for PQD 2, 3, and 4 including the expected evolution according to the GDJ model. PQD 3 shows only minor broadening on the timescales investigated suggesting photon emission with high spectral stability.

**Materials and Methods:**

Chemicals

Lead(II) acetate trihydrate (>=99%, Aldrich), oleic acid (OA, 90%, Aldrich), Br$_2$ (99%, Aldrich), 1-octadecene (ODE, 90%, Aldrich), toluene (HPLC, Aldrich), 3-(N,N-dimethyloctadecylammonio)propanesulfonate (>99%, Aldrich), Cesium carbonate (99%, Fluorochem), trioctyl phosphine (TOP, 97%, STREM) and ethyl acetate (HPLC, Fischer) were used as received.

Cesium-Oleate Synthesis

1.628 g of Cs$_2$CO$_3$ (10 mmol, 1 eq Cs) and 5 mL of oleic acid (16 mmol, 0.8 eq) were evacuated in a three-neck flask along with 20 mL of ODE at room temperature until the first gas evolution subsides and then further evacuated at 25-120 °C for 1 hour. This yields a 0.4 M solution of Cs-oleate in ODE. The solution turns solid when cooled to room temperature and was stored under argon and heated before use.

Lead(II)-Oleate Synthesis

4.6066 g of lead (II) acetate trihydrate (12 mmol, 1 eq) and 7.6 mL of oleic acid (24 mmol, 2 eq) were evacuated in a three-neck flask along with 16.4 mL of ODE at room temperature until the first gas evolution subsides and then further evacuated at 25-120 °C for 1 hour. This yields a 0.5 M solution of Pb-oleate in ODE. The solution turns solid when cooled to room temperature and was stored under argon and heated before use.

TOPBr$_2$ Synthesis

The solution was prepared in a Schlenk flask under argon by mixing 6 mL TOP (13 mmol, 1 eq) with 0.6 mL Br$_2$ (11.5 mmol, 0.88 eq). As soon as the reaction was completed the TOPBr$_2$-adduct was dissolved in toluene (18.7 mL) to create a 0.5 M (light yellow) solution.

Perovskite Nanocrystal Synthesis

Zwitterionic-ligand-capped CsPbBr$_3$ NCs were synthesized according to the method previously published in (ACS Energy Lett. 2018, 3, 641−646). Lead oleate (5 mL, 0.5 M in octadecene, Cs-oleate (4 mL, 0.4 M in octadecene) and N,N-(dimethyloctadecylammonio)-propanesulfonate (commercial zwitterionic detergent, 0.2145 g, 0.5 mmol) and 50 mL of octadecene were added to a 100 mL three neck flask and heated to 120°C under vacuum, where the atmosphere was changed to nitrogen. The mixture was then heated to 180°C, followed by the injection of trioctylphosphine-bromine adduct (TOPBr$_2$ , 5 mL, 0.5M in toluene). The reaction flask was immediately cooled to room temperature with an ice-bath. Two fractions were separated by centrifugation at 30'000g (g being the unit of the Earth gravity) for 10 minutes. The supernatant fraction was discarded and the precipitate fraction was redispersed in 20 mL of toluene. These NCs were purified by repetitive (3 times) addition of ethylacetate (two volumetric equivalents) followed by centrifugation at 30'000g for 1 minute and subsequent re-dispersion in toluene (10 mL after first precipitation, 5 mL after subsequent precipitations).

Ensemble Characterization

Powder X-Ray Diffraction patterns (XRD) patterns were collected with STOE STADI P powder diffractometer, operating in transmission mode. A germanium monochromator, Cu Kα1 irradiation and a silicon strip detector (Dectris Mythen) were used. Standard Transmission Electron Microscopy (TEM) images were collected using Hitachi HT7700 microscope operated at 100 kV. TEM images were processed using Image J (or new Fiji).*(38)* High-resolution TEM images (Figure 1) were acquired through a JEOL 2010 Advanced High Performance TEM operating at 200 kV using a lanthanum hexaboride cathode. Samples were prepared by drop casting NC solutions in toluene onto 400 mesh copper grids supporting an amorphous carbon film.

Optical absorption and photoluminescence (PL) and absolute solution quantum yield (QY) measurements
Optical absorption UV-Vis absorption spectra for colloidal solutions were collected using a Jasco V670 spectrometer in transmission mode. A Fluorolog iHR 320 Horiba Jobin Yvon spectrofluorometer equipped with a PMT detector was used to acquire steady-state PL spectra from solutions. QYs from green PQDs dispersions were estimated using fluorescein as a reference, according to the methods suggested by IUPAC *(39)*.

Single Quantum Dot Characterization
Samples for single molecule spectroscopy were prepared by diluting the stock-solution of PQDs in a PMMA (3 % by weight in toluene) solution, and spincoating (5000 rpm, 1 min) the solution on quartz slides (MTI, optical grade). The samples were cooled to liquid helium temperatures with a closed cycle liquid helium cryostat (Cryostation, Montana Instruments). Single particle spectroscopy was performed using a home-built confocal fluorescence microscope. Single PQDs were excited with a 488nm CW laser (Coherent Sapphire) for the recording of emission spectra and the PCFS experiments. The single particle emission was filtered spatially (telescope (8 cm focal length with a 30 μm pinhole) and spectrally (dichroic notch (488nm, Edmund Optics) and longpass filters (500nm, Thorlabs)). Single particle spectra were recorded by diffracting the mission in a monochromator (Acton 2050, Princeton Instruments, 600 g/mm) and detection with a cooled EMCCD camera (ProEM512B, Princeton Instruments). For the PL lifetime and g2 measurements, single nanocrystals were excited non-resonantly with either a frequency doubled (404 nm Ti-Sapphire crystal) pulsed laser (Coherent Mira 900F, 808nm 80MHz) with a nominal pulse width of 30fs or a TOPTICA Systems FemtoFiber Pro Vis (488nm, 80MHz, <1ps pulsewidth). The emission was collected in time-tagged mode (T3 mode, Picoquant Hydraharp) using a fast photodiode as the trigger. Photon-detection was performed with a fast APD (Micro Photon Devices) with a nominal timing resolution of FWHM ~50ps. To observe the strong anti-bunching, the sharp zero phonon line of the emission was spectrally isolated with tunable long- and short-pass filters (Semrock Versachrome TLP01-561 and TSP01-561).

Photon-Correlation-Fourier-Spectroscopy
PCFS is described in detail experimentally and theoretically previously*(19, 33)*. The single emitter photon-stream is directed into a variable path-length-difference Michelson interferometer (Newport Stage), and detected with two APDs (AQRH-16, Excelitas) in conjunction with picosecond photon-timing card (Picoquant GmbH, Hydraharp). The path-length difference is adjusted to discrete positions within one half of the single emitter interferogram (80-120 positions in total) and dithered with a triangle waveform over 2000 nm and a frequency of 0.05 Hz. The photon-stream at each stage-position was collected for 1 or 2 min to achieve sufficient S/N at short photon-lag times. The PCFS data analysis is performed by calculating the second-order intensity cross-correlation of photons detected with the two APDs and the auto-correlation of their sum-signal. The PCFS interferogram is mirrored around the white-fringe stage-position and Fourier transformation gives the spectral correlation.

# Supplementary Information for:
# Coherent Single Photon Emission from Colloidal Lead Halide Perovskite Quantum Dots

Hendrik Utzat[1], Weiwei Sun[1], Alexander E.K. Kaplan[1], Franziska Krieg[2,3], Matthias Ginterseder[1], Boris Spokoyny[1], Nathan D. Klein[1], Katherine E. Shulenberger[1], Collin F. Perkinson[1], Maksym V. Kovalenko[2,3], and Moungi G. Bawendi[1]*

[1] Department of Chemistry, Massachusetts Institute of Technology, 77 Massachusetts Avenue Cambridge, MA 02139.

[2] Institute of Inorganic Chemistry, Department of Chemistry and Applied Bioscience, ETH Zurich, 8093 Zurich, Switzerland.

[3] Laboratory for Thin Films and Photovoltaics, Empa − Swiss Federal Laboratories for Materials Science and Technology, CH-8600 Dübendorf, Switzerland.

*Correspondence to: mgb@mit.edu

Equations and Notation Describing PCFS

In PCFS, we collect the photon-stream at the output arms of a Michelson interferometer while dithering the path-length difference $\delta$ over multiple fringes around a center position $\delta_0$. The spectral information of the emitter is encoded in the intensity cross-correlation function $g_\times^{(2)}$ defined as

$$g_\times^{(2)}(\delta_0) = \frac{\langle I_a(t) I_b(t+\tau) \rangle}{\langle I_a(t) \rangle \langle I_b(t+\tau) \rangle} \qquad (1)$$

where $\langle \dots \rangle$ denotes the time average. In this manuscript, we use PCFS on time-scales larger than photon anti-bunching timescales, and the intensity traces due to single-photon interference can be described classically with the well-known result

$$I_{a,b}(t) = \frac{1}{2} I(t) \big[ 1 \pm \text{FT}\{s(\omega,t)\}_{\delta(t)} \big] \qquad (2)$$

where we have introduced the time-dependent single emitter spectrum $s(\omega, t)$ and the time-dependent intensity of the single emitter $I(t)$. Introducing (2) in (1), we obtain:

$$g_\times^{(2)}(\delta_0, \tau) = \frac{\langle I(t) I(t+\tau) \big[1 + \text{FT}\{s(\omega,t)\}_{\delta(t)}\big]\big[1 - \text{FT}\{s(\omega,t)\}_{\delta(t+\tau)}\big]\rangle}{\langle I(t)\big[1 + \text{FT}\{s(\omega,t)\}_{\delta(t)}\big]\rangle \langle I(t+\tau)\big[1 - \text{FT}\{s(\omega,t)\}_{\delta(t+\tau)}\big]\rangle} \qquad (3)$$

Equation (3) can be simplified by expansion and by making use of the fact that the time-average $\langle \ldots \rangle$ over intensity terms of the form $\text{FT}\{s(\omega,t)\}_{\delta(t)}$ or $\text{FT}\{s(\omega,t)\}_{\delta(t+\tau)}$ vanish as we ensure experimentally that for each recorded correlation function $\int_0^t \delta(\tau)\, d\tau = \delta_0$:

$$g_\times^{(2)}(\delta_0, \tau) = \frac{\langle I(t)I(t+\tau)\rangle}{\langle I(t)\rangle\langle I(t+\tau)\rangle}\left(1 - \langle \text{FT}\{s(\omega,t)\}_{\delta(t)} \times \text{FT}\{s(\omega,t)\}_{\delta(t+\tau)}\rangle\right) \quad (4)$$

where the first term is just the intensity auto-correlation function $g_{||}^2$.
We then introduce $G^{(2)}(\delta, \tau)$ defined as:

$$G^{(2)}(\delta, \tau) = 1 - \frac{g_\times^{(2)}(\delta_0, \tau)}{g_{||}^2(\tau)} = \langle \text{FT}\{s(\omega,t)\}_{\delta(t)} \times \text{FT}\{s(\omega,t)\}_{\delta(t+\tau)}\rangle \quad (5)$$

where we have introduced the spectral correlation $p(\zeta, \tau)$, the auto-correlation of the spectrum compiled from photon-pairs separated by $\tau$:

$$p(\zeta, \tau) = \left\langle \int_{-\infty}^{\infty} s(\omega, t) s(\omega + \zeta, t + \tau)\, d\omega \right\rangle \quad (6)$$

The Gaussian Discrete Spectral Jump Model

We describe the spectral diffusion observed herein as a spectrum $s(\omega, t)$ undergoing discrete jumps to spectral positions defined by a Gaussian probability distribution and a time constant $t_c$. The physical interpretation for this model is that random configurations of the dielectric field change discretely inducing a dynamic Stark shift of the homogeneous spectrum. Considering a first order process for the jump dynamics with a single jump rate $k = 1/t_c$, the probability of at least one random spectral jump having occurred is related to the characteristic first order decaying exponential function

$$p_{\text{jump}} = 1 - e^{-k\tau}.$$

With our assumption that spectral jumps are uncorrelated and confined by a Gaussian envelope, the total spectral correlation $p(\zeta, \tau) = <\int_{-\infty}^{\infty} s(\omega, t) s(\omega + \zeta, t + \tau) d\omega>$ can be parsed into components that are weighted by the jump probability according to

$$p(\zeta, \tau) = \left(1 - p_{\text{jump}}(\tau)\right) p_{\text{homogeneous}}(\zeta) + p_{\text{jump}}(\tau) p_{\text{envelope}}(\zeta),$$

where

$$p_{\text{envelope}}(\zeta) = \int_{-\infty}^{\infty} e^{-\frac{(\omega-\omega_0)^2}{2\sigma^2}} e^{-\frac{(\omega-\omega_0+\zeta)^2}{2\sigma^2}} d\omega$$

is just the auto-correlation of the envelope confining the spectral jumps. Intuitively, because PCFS averages over all photon-pairs with a given temporal separation $\tau$ and jumps are uncorrelated in time, probability density is transferred from the undiffused spectral correlation with $\tau \ll t_c$ to the diffused spectral correlation at larger $\tau$.

To model the evolution of the spectral correlation due to spectral diffusion as shown in Figure 4 of the main text, we use the fit to the small $\tau$ (100 $\mu$s) spectral correlation as $p_{\text{homogeneous}}$. The linear combination of $p_{\text{homogeneous}}(\zeta)$ and $p_{\text{envelope}}(\zeta)$ with the $\tau$-dependent weights is used to extract the FWHM evolution numerically (Figure S7). The width of the Gaussian envelope $\sigma$ and the characteristic jump time $t_c$ are fit parameters to describe the data in Figure 4 of the main text.

Determination of the low-temperature emission quantum yield of PQDs

The low-temperature photoluminescence emission quantum yield (PLQY) was determined by monitoring the photoluminescence emission intensity of a thin-film of PQDs and normalizing it by the absolute room temperature emission quantum yield of the same films, following the method described by de Mello, Wittmann, and Friend*(33)*. For this, we used a 5 mW, 405 nm diode excitation laser (Thorlabs LDM405) and a calibrated integrating sphere (Labsphere), fiber-coupled to a spectrometer (Ocean Optics Flame-NIR). The spectrometer was wavelength calibrated using a mercury-argon lamp (Ocean Optics HG-1) and intensity calibrated using an absolute irradiance-calibrated tungsten halogen lamp (Ocean Optics HL-3P-CAL).

A sample was prepared by drop-casting $CsPbBr_3$ quantum dots from toluene onto a quartz substrate (MTI) and drying overnight. The PLQY of the sample was measured six times at 298 K, yielding a value of (70.8 ± 1.5)%. To obtain the low-temperature PLQY, the sample was then loaded into a helium cryostat (Janis VPF-100) mounted in a UV-Vis spectrofluorometer (Horiba FluoroMax-3), and photoluminescence spectra were measured with constant 405 nm excitation at several temperatures between 298 K and 5 K.

Fig. S8 (B) presents emission spectra of the quantum dot film measured at different temperatures. The spectra were integrated over the emission peak to determine total relative photoluminescence intensity. An increase in photoluminescence intensity by a factor of 1.35 ± 0.04 was measured at 150 K relative to 298 K. No further brightening was observed as the sample was further cooled to 5 K. Meanwhile, the optical absorption of the quantum dots is expected to remain nearly constant over temperatures between 298 K and 5 K. From the room temperature PLQY and emission enhancement factor upon cooling the sample, the PLQY of the $CsPbBr_3$ quantum dots at 5K was determined to be (95.6 ± 4.2)%.

**Fig. S1.**

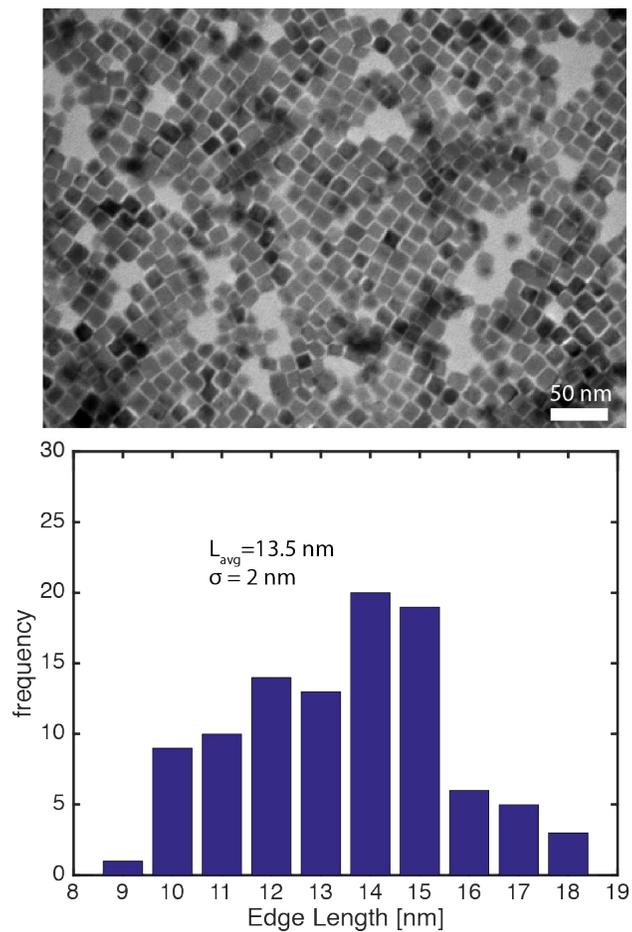

Additional representative TEM image of the Br-PQDs used in this study. The particles exhibit a cubic shape with an average edge length of 13.5 nm and a standard deviation of 2 nm as can be seen in the corresponding edge-length histogram.

**Fig. S2.**

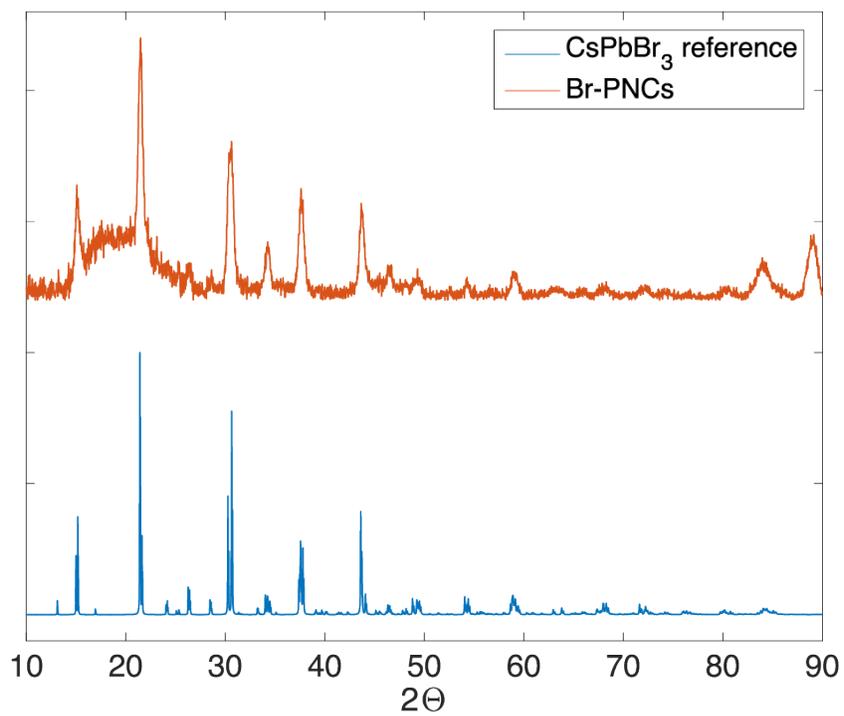

Powder X-Ray diffraction pattern of Br-PQDs similar to the ones used in the study. The PQDs show an orthorhombic crystal structure (P*nma*) as reported previously. We show the bulk CsPbBr$_3$ P*nma* space-group reference diffraction pattern in blue.

**Fig. S3.**

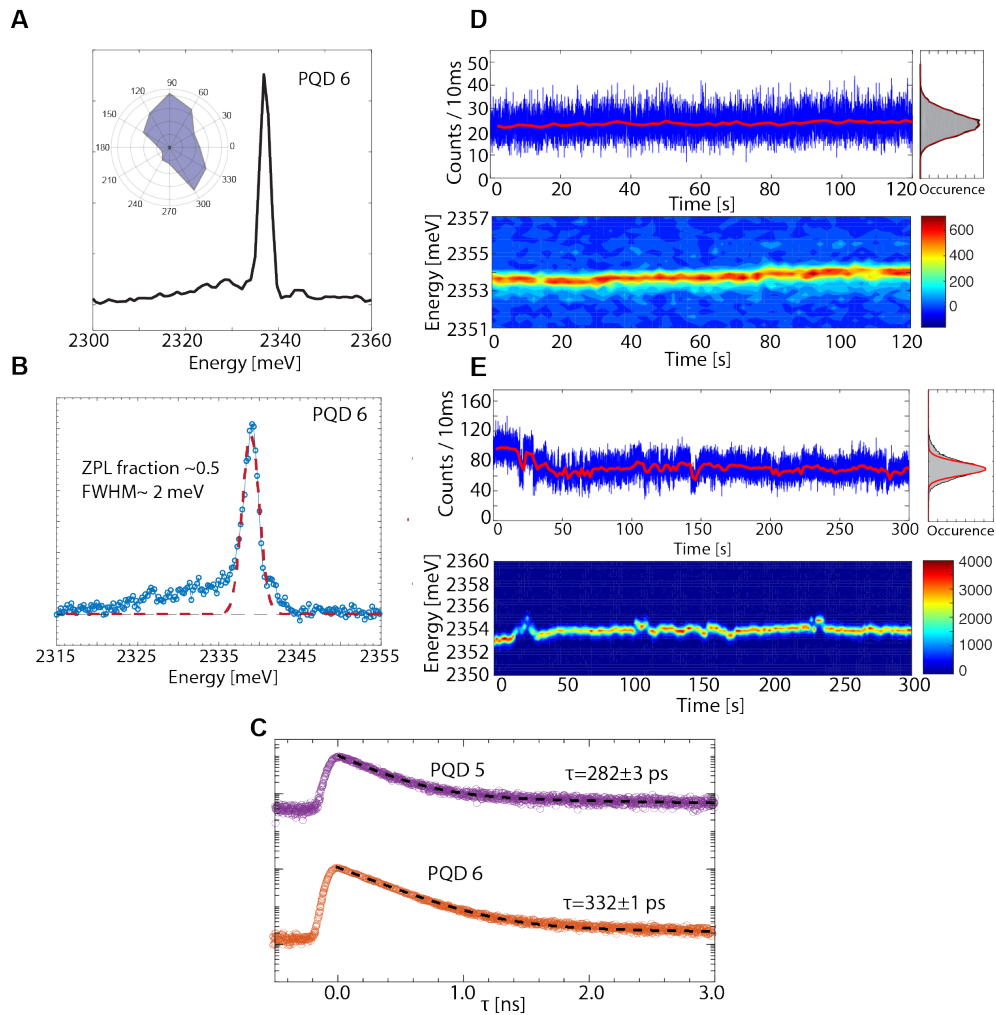

Additional basic single PQD characterization spectra for PQD 6 with ~500μeV and ~150 μeV spectrometer resolution (**A**, **B**). A broad background, likely due to fast relaxation within the fine-structure or a broad acoustic phonon feature, can be discerned. Additional PL lifetimes with fast emission are shown in (**C**). Additional representative spectral and intensity traces for two PQDs are shown in (**D**, **E**). Some PQDs show absence of fluorescence intermittency over the course of minutes (**D**). Typical excitation intensities of 20-50 W/cm$^2$ (488nm) were used to acquire these traces. The spectral time series shows slow spectral jump dynamics not exceeding ~ 1 meV. Large spectral jumps in colloidal QDs are typically associated with trion formation. As the trion binding energy of our particles is ~ 16 meV (Fig. S5), consistent with previously reported values*(17)*, trion formation is unlikely the origin of the slow spectral dynamics. The observed slow spectral dynamics is likely due to discrete changes in the PQD dielectric ligand-environment*(5)*.

**Fig. S4.**

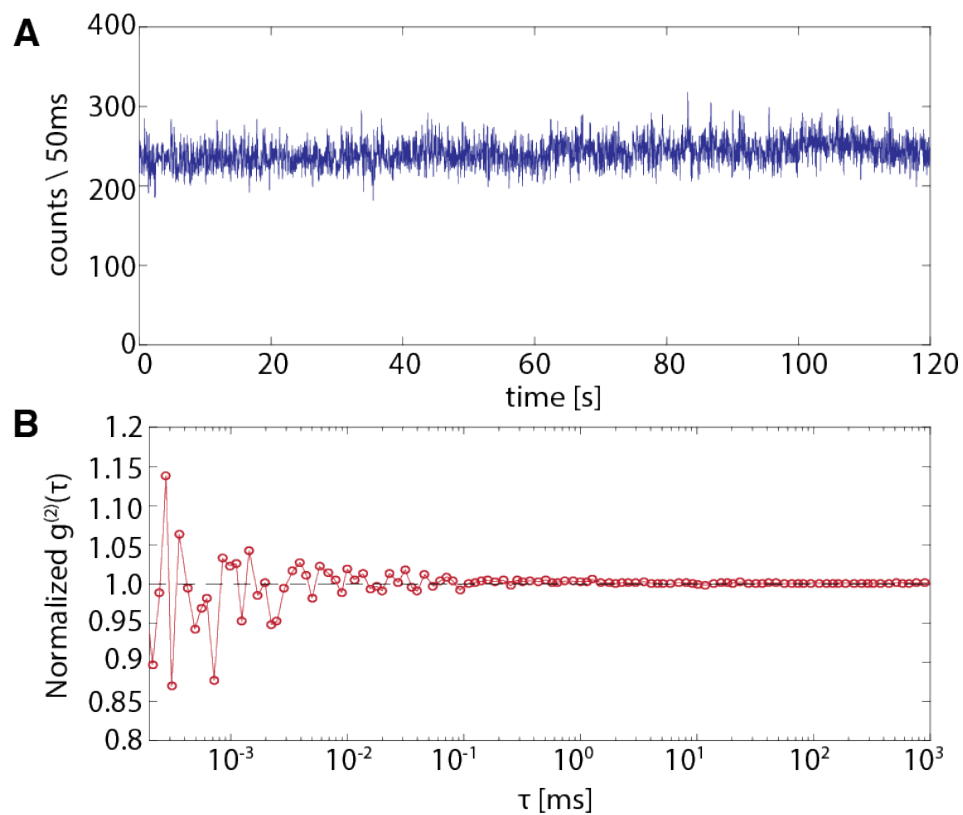

Intensity trace of a single PQD showing absence of fluorescence intermittency over the course of minutes under cw excitation (488nm, 100W/cm$^2$)(**A**). The trace is detector shot-noise limited. We show that the emission of PQDs can be near Poissonian on timescales from μs to s by plotting the second order intensity correlation function between two detectors as shown in (**B**). $g^{(2)}(\tau)$ is ~1 on these timescales indicating near random temporal photon distributions and the absence of fast blinking processes.

**Fig. S5.**

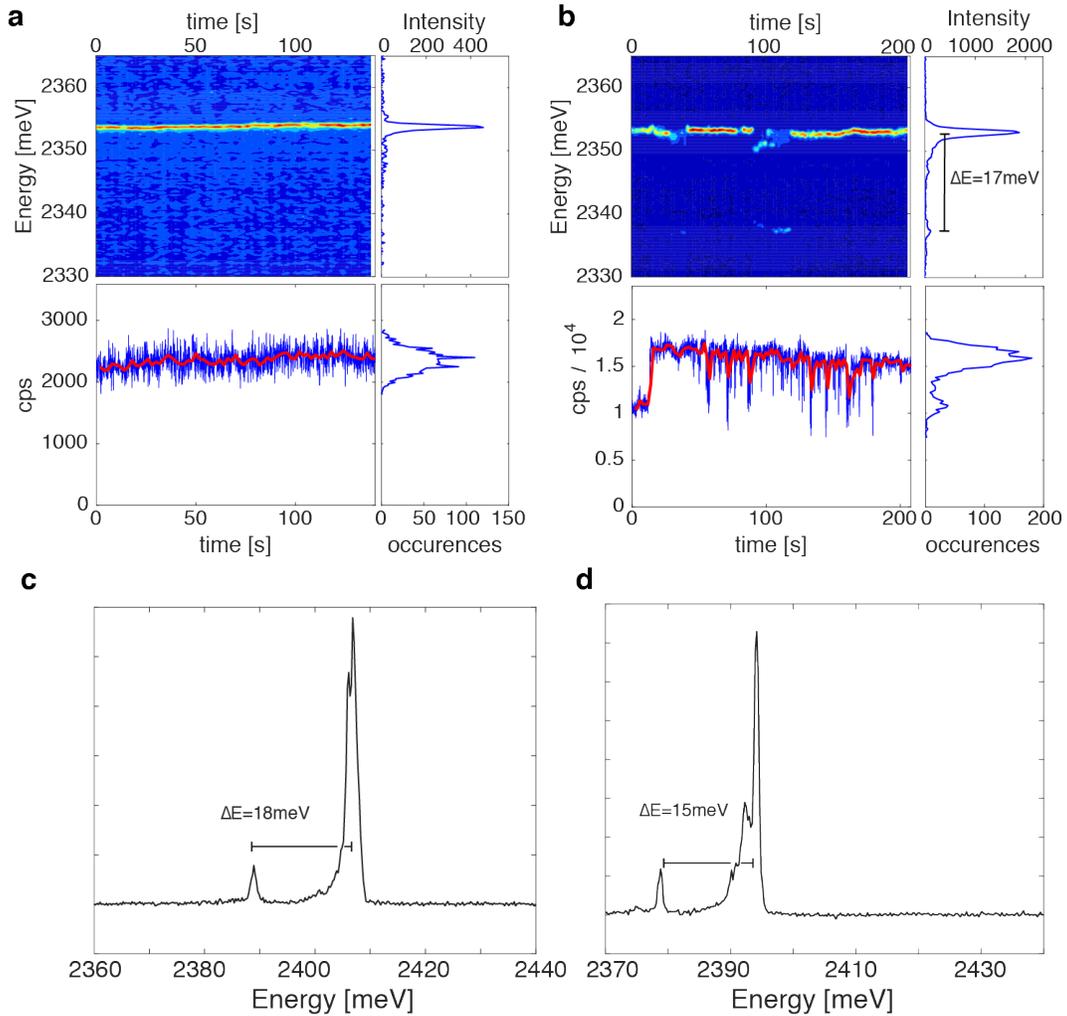

Spectral time-series and intensity trace acquired at typical excitation intensities of 50 W/cm$^2$ (**A**) and at 500 W/cm$^2$ (**B**). Some PQDs show absence of fluorescence intermittency over the course of minutes in their intensity traces and no spectral jumping as indicated by a single sharp peak in the emission spectrum. At higher intensities, fluorescence intermittency and spectral jumping over ~ 17 meV is observed. We assign the lower energy peak to the trion. Additional spectra of single PQDs under high excitation flux with trion peaks offset by 18 and 15 meV are shown in (**C**) and (**D**). We performed all single emitter experiments at low excitation intensities of ~ 50 W/cm$^2$ to avoid contributions from multiexciton states and photo-charging (trion formation).

**Fig. S6.**

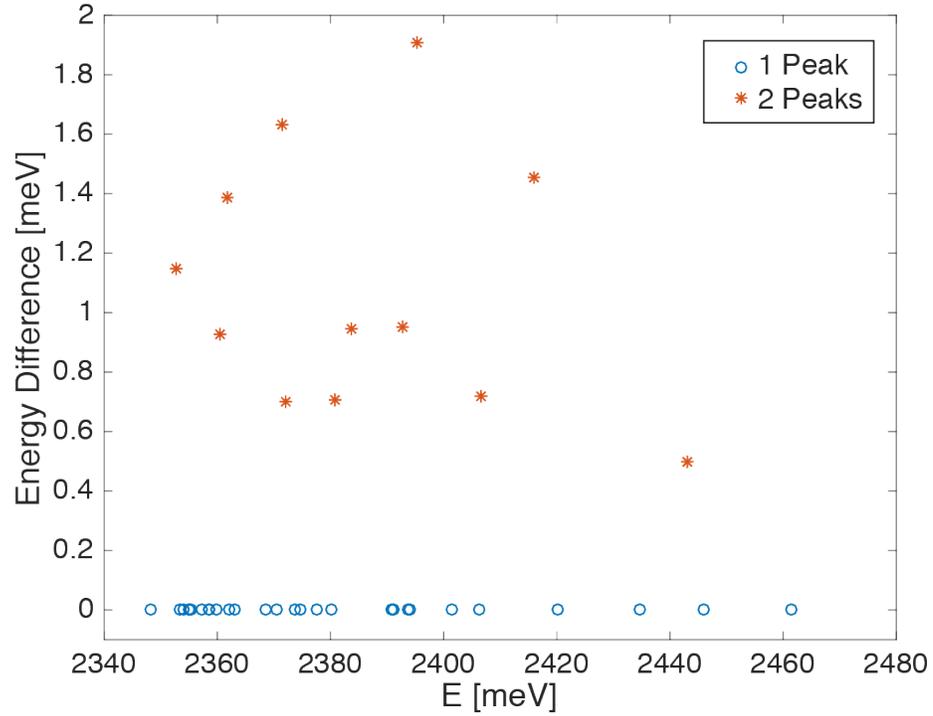

Distribution of center emission energies of PQDs with one and two peaks resolved with diffraction-grating-based spectroscopy. The resolution limit of our spectrometer is ~500 µeV. PQDs with fine-structure splittings significantly smaller will appear to have one peak. No clear correlation between fine-structure splitting and center emission energy can be discerned.

**Fig. S7.**

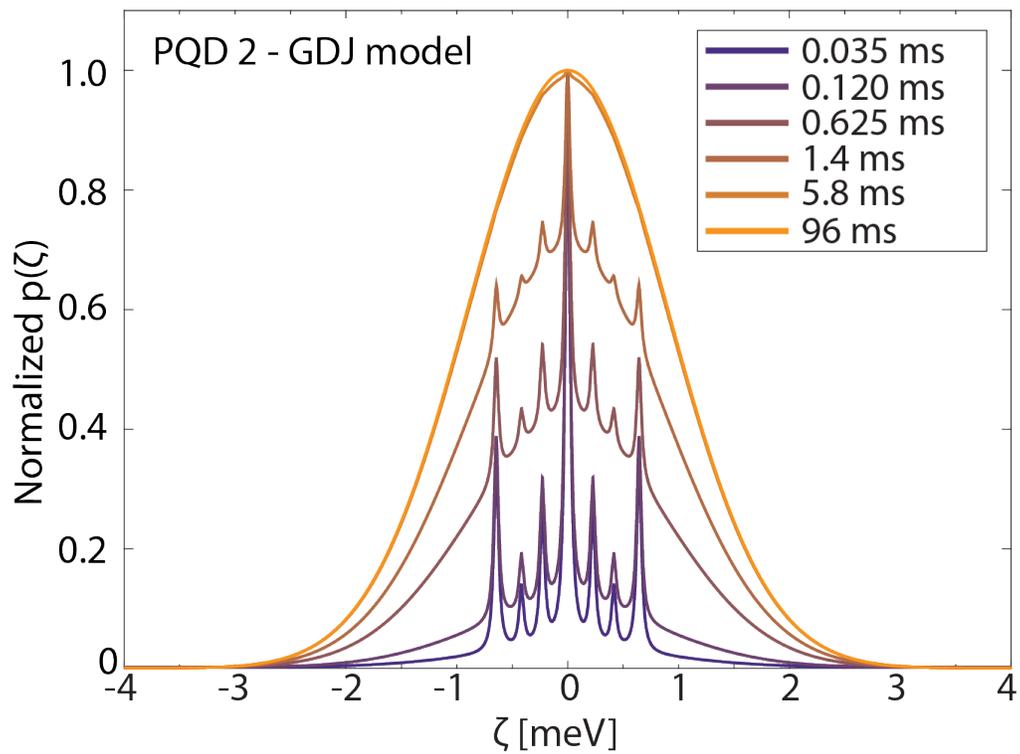

Evolution of the spectral correlation $p(\zeta, \tau)$ with $\tau$ simulated for PQD 2 (Fig. 4 **A**) with three observable fine-structure states and a characteristic jump time of $\tau$=0.84ms. The temporal progression is described by a linear combination of the undiffused spectral correlation extracted from the fit in Fig. 3 of the main text and a broad Gaussian feature defined by the envelope of discrete spectral jumps.

**Fig S8.**

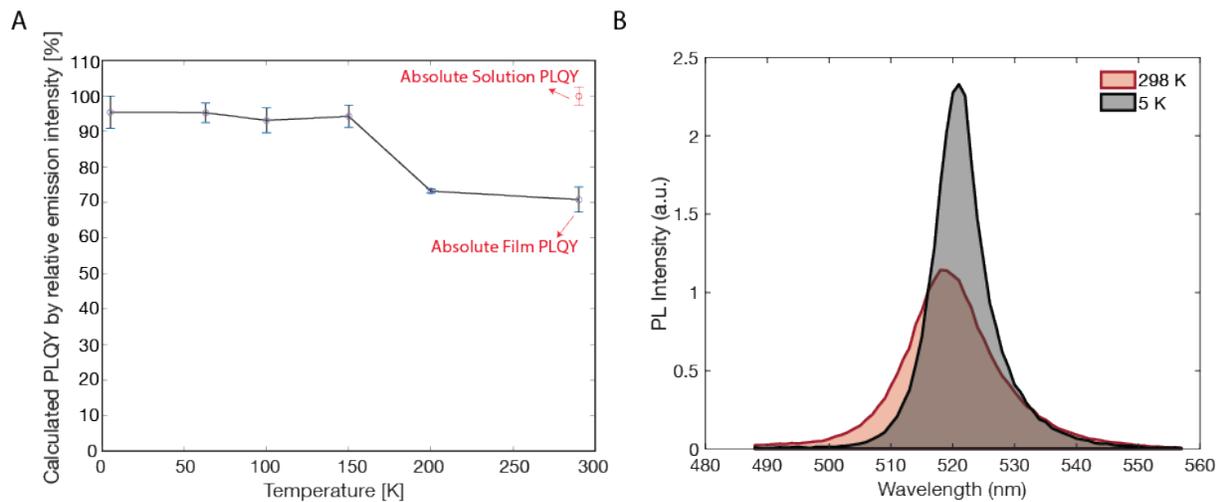

Evolution of the ensemble photoluminescence emission quantum yield (PLQY) for bromide PQDs (**A**). The room-temperature solution PLQY was determined to be near unity. Upon drop-casting a dilute thin-film of PQDs, the PLQY slightly reduced to ~70%. This is likely due to exciton self-quenching or presence of more non-radiative channels in the solid-state environment compared to solution. Upon cooling the thin-film to 5K, the emission intensity increases (**B**) to recover a near unity (95±4%) PLQY at low temperatures.